\documentclass[draft,grl]{agutex2015}

\usepackage{graphicx}

 \setkeys{Gin}{draft=false}

\authorrunninghead{OBENBERGER ET AL.}

\titlerunninghead{Altitudes of Meteor Radio Afterglows }

\authoraddr{Corresponding author: K.S. Obenberger,
(obenken@icloud.com)}

\begin{document}

\title{Altitudinal dependence of meteor radio afterglows measured via optical counterparts}

\authors{K. S. Obenberger\altaffilmark{1}, J. M. Holmes\altaffilmark{1}, J.D. Dowell\altaffilmark{2}, F. K. Schinzel\altaffilmark{2}, K. Stovall,\altaffilmark{2}, E. K. Sutton\altaffilmark{1}, and G. B. Taylor\altaffilmark{2}}

\altaffiltext{1}{Space Vehicles Directorate, Air Force Research Laboratory, Kirtland AFB, New Mexico, USA}
\altaffiltext{2}{University of New Mexico, Albuquerque, NM, USA}

 \keypoints{\item We have measured the positions of 44 meteor radio afterglows
 \item Meteor trails above 90 km are strongly preferred to produce radio emission
 \item This dependence agrees with the plasma emission hypothesis}

\begin{abstract}
Utilizing the all-sky imaging capabilities of the LWA1 radio telescope along with a host of all-sky optical cameras, we have now observed 44 optical meteor counterparts to radio afterglows. Combining these observations we have determined the geographic positions of all 44 afterglows. Comparing the number of radio detections as a function of altitude above sea level to the number of expected bright meteors we find a strong altitudinal dependence characterized by a cutoff below $\sim$ 90 km, below which no radio emission occurs, despite the fact that many of the observed optical meteors penetrated well below this altitude. This cutoff suggests that wave damping from electron collisions is an important factor for the evolution of radio afterglows, which agrees with the hypothesis that the emission is the result of electron plasma wave emission.
\end{abstract}

\begin{article}

\section{Introduction}\label{Intro}

Recently \citet{Obenberger14} discovered that bright meteors will occasionally produce a  radio afterglow at the high frequency (HF; 3 - 30 MHz) and very high frequency (VHF; 30 - 300 MHz) radio bands. Since then, afterglows have been regularly observed using the all-sky imaging capabilities of the first station of the Long Wavelength Array (LWA1), a 10 to 88 MHz radio telescope comprised of 256 dual polarization dipole antennas spread over a diameter of 100 meters \citep{Ellingson13}. The afterglows have little to no polarization, last from tens to hundreds of seconds, and have a broad spectrum, which appears to follow a steep power law, getting brighter at lower frequencies \citep{Obenberger15b,Obenberger16}. An upper frequency cutoff has not been detected in any event, but the radiated power drops below the LWA1 sensitivity above 60 MHz, for most events.

We believe the emission to be intrinsic to meteor events and not the result of reflection. It is well established that meteors create ionization trails that can reflect man-made or or natural radio emission. Reflected signals are generally narrowband, polarized, and irregular, which is in contrast to the properties of the radio afterglows that we observe with the LWA1. There are no known radio sources that are bright enough and have sufficient bandwidth to produce reflected signals similar to our observations.

With nearly 20,000 hours of data collected between April 2012 and April 2016, 154 radio transients have been detected by the LWA1, the majority of which appear to be radio afterglows from meteors. \citet{Obenberger15b} proposed the hypothesis that the afterglows could be the result of radiated electron plasma waves, occurring across a range of plasma frequencies found in the turbulent trail. This hypothesis was based on the fact that the emission is broadband, and appears to trace the expected range of plasma frequencies of the trail, where the plasma frequency is related to the electron density, $n_{e}$, by,
\begin{equation}
\omega_{pe} = \sqrt{\frac{n_{e} e^{2}}{m \epsilon_{0}}} 
\end{equation}
where, $e$ is the electric charge, $m$ is the mass of the electron, and $\epsilon_{0}$ is the permittivity of free space.

The electron plasma wave hypothesis has several challenges. Electron collisions with neutrals and ions would damp the waves much faster than the long timescales over which the afterglows are observed, requiring energy to be continuously injected into the trail in order to drive waves. Furthermore, electrostatic electron plasma waves would need to somehow be converted into electromagnetic waves, but steep density gradients within meteor trails would certainly aid the process. This is likely analogous to stimulated electromagnetic emissions (SEE) observed in the ionospheric F-region, where Langmuir waves are driven near the center frequency of high power transmitters and are then converted to electromagnetic waves \citep{Leyser01}.

If radio afterglows are indeed caused by electron plasma waves, there should be an altitudinal dependence on the occurrence and brightness features. Electron plasma waves can only occur at altitudes where the electron collision frequency with both neutrals and ions is lower than the plasma frequency. More importantly, collisions would act as a damping agent to the wave growth.

Currently, the best way to measure the altitudes of radio afterglows is by triangulating the position of optical  counterparts. This is possible using the many all-sky fireball-searching cameras located in New Mexico or Arizona that share visible sky with the LWA1. The NASA All-Sky Fireball Network \citep{Cooke12} operates two cameras in southeast NM and three in south central AZ. Likewise the Sky Sentinel LLC (http://goskysentinel.com), operates two cameras near Albuquerque, NM. The authors of this paper operate one camera at Sevilleta National Wildlife Refuge, NM. Most of the cameras are all well over 100 km away from the LWA1, meaning that many events that occur above the LWA1 are near or below the observable horizon. Furthermore, light pollution (man-made and from the Moon) along with cloudy nights prevent many LWA1 events from being seen by the cameras.

Despite these difficulties, between April 2012 and June 2016 we detected 44 meteor radio afterglows with measured optical counterparts.  Statistical analysis of these events are presented in this paper.

\section{Observations}

Radio afterglows are easily found using the Prototype All-Sky Imager (PASI), a back end to the LWA1 telescope. The imaging and meteor afterglow search processes are described in \citet{Obenberger15a} and \citet{Obenberger14}. Once a meteor afterglow is identified we search for an optical counterpart as viewed from any of the all-sky video (30 FPS) cameras operated in New Mexico and in Arizona. 

\begin{figure}
	\centering
	\includegraphics[width = 3.5in]{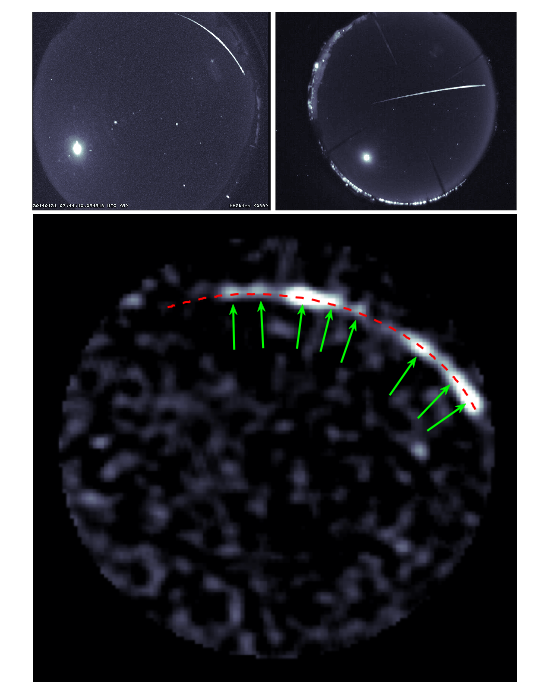}
	\caption{(Top) Integrated video still of a meteor from both the NASA/MEO camera in Mayhill, NM (left) and the Sky Sentinel camera in Albuquerque, NM (right). (Bottom) All-sky radio image of the meteor radio afterglow, recorded with PASI at 38 MHz and averaged over 15 seconds. The optical/optical triangulated meteor path, translated to the image coordinates of the LWA, is shown as a red dotted line. Eight individual radio brightness components are highlighted with green arrows. The angular error of the optical/optical triangulated meteor path is much less than the pixel size of the radio image. For each image North is up, and West is right}
	\label{image}
\end{figure}

The two southern NM cameras and the three in Arizona are operated by the NASA Meteoroid Environment Office (MEO, http://fireballs.ndc.nasa.gov) and use the All Sky and Guided Automatic Realtime Detection (ASGARD) Software \citep{Weryk08,Brown10}. The other two New Mexico Cameras are located in Albuquerque, were built by Sandia National Laboratory, are operated by Sky Sentinel LLC and use WSentinel software for meteor identification. Figure \ref{image} shows the integrated video stills of the same meteor as viewed from the NASA/MEO camera near Mayhill, NM and the Sky Sentinel Camera in Albuquerque, NM as well as the PASI image of the radio afterglow.



The authors of this paper operate another camera (APOGEE U2000 and 2.7mm f/1.8 Fujinon fisheye lens)  located at the LWA station at the Sevilleta National Wildlife Refuge (LWA-SV). Unlike the Sky Sentinel and NASA/MEO cameras, this camera does not operate at video rates. Rather it takes 5 second exposures every $\sim$ 7.5 seconds. The data is still useful since we can triangulate the observed meteor streak with the point-by-point measurements from single station video cameras.

The 5 second images are geometrically calibrated using the open source AIDA\_tools software package for Matlab, whereby image pixels are mapped to elevation and azimuth sky angles. Roughly 200 stars can be fitted to better than 1 pixel difference down to about 10 degrees elevation angle.



%

\section{Afterglow Position Calibration}

While ASGARD software performs the triangulation between all NASA/MEO cameras, combinations involving at least one Sky Sentinel camera, the Sevilleta camera, or the LWA1 use a triangulation function written in MATLAB by the authors of this paper. This function uses the time series of azimuth and elevation angles derived from calibrated pixel locations. The pixel locations are found semi-manually from the recorded videos, using an image subtraction function also written in MATLAB. In the case where two video cameras are used, corresponding points from each video are determined by both the time stamp and assuring that brightness features match up in time. In the case where a video and 5 second snapshot are used, corresponding points are determined by matching brightness features and making assumptions about the angular size comparison, namely that the meteor is at a range of $h/ \sin{(\phi)}$, where $h$, the altitude of the event above sea level, is assumed to be 100 km and $\phi$ is the elevation angle.


\subsection{Optical/Optical Triangulation}

In most situations where an optical counterpart is observed, it is seen by at least two cameras, allowing for an optical/optical triangulation. The triangulated path of an optical counterpart trail can then be used to calibrate the position of the afterglow seen by the LWA1. This is accomplished by comparing LWA1 measured coordinates of the radio afterglow to a locus of points along the optical/optical triangulated trail. The location of the radio afterglow is determined to be the position along the trail that best fits the LWA1 measured coordinates. For instance, Figure \ref{image} shows a radio afterglow with multiple radio components and the optical/optical triangulated trail (red dotted line). The position of each radio component is determined to be the part of the optically triangulated trail closest to the peak of that component.

The all-sky cameras used to triangulate the meteors have angular errors around $\pm 0.2^{\circ}$ \citep{Brown10}, which is much smaller than the LWA1 angular error of $\pm 2^{\circ}$. The LWA1 contribution, therefore, dominates. The radio afterglow shown Figure \ref{image} highlights this fact, where the error in the optical/optical triangulated meteor trail is smaller than the pixel size in the radio image. The LWA1 error, on the other hand, is about the size of a pixel. The LWA1 $2^{\circ}$ error estimate was originally derived from celestial sources in the images. We note that the residual differences between the measured coordinates of radio afterglows and the final positions fit along the counterpart trails agree with the $2^{\circ}$ estimate.

The $2^{\circ}$ angular error can be converted to more meaningful positional units of kilometers using $R\times \sin{(2^{\circ})}$, where $R$ is the range to the meteor from the LWA1. For all events evaluated in this study the median range was 120 km, with the closest being 90 km and furthest being 260 km. These ranges translate to a median error of $\pm$4 km, a minimum error of $\pm$3 km, and maximum error of $\pm$9 km.

\subsection{Optical/Radio Triangulation}
Optical/optical triangulations are most desirable since they allow for the determination of the position, velocity, and optical magnitude for the entire meteor path. However, we are also able to triangulate measurements from a single optical camera with measurements from the LWA1. Due to poor LWA1 resolution, along with the fact that the radio afterglows neither trace the entire path of the meteor nor the brightness profile,  we are not able to compare features to determine the part(s) of the meteor path from which the radio afterglows are emanating. Instead, we triangulate the positions of the radio components with the entire optical path of the meteor. This results in two sets of geographical positions, one calculated from the LWA1's perspective and one from the camera's perspective. The location where the two sets of positions intersect (or most closely match), is determined to be the actual location of the radio afterglow. This method was compared to the optical/optical method and the two give results within the LWA1 angular error.

\section{Results}

Currently 154 transients have been found in PASI images captured since 2012. Of these events, 89 have occurred between sunset and sunrise, and of these, 32 have been seen by at least two all-sky cameras allowing for optical/optical triangulation. An additional 12 were seen by only one camera, allowing for only a optical/radio triangulation. This leaves 45 events that were not seen by any optical camera. However, it should be noted that 20 of these unseen events occurred on nights that were reported as overcast at one or more camera locations. 

\subsection{Radio Energy Dependence on Optical Magnitude }

Since we do not have a model for the full radio spectrum we cannot compute the total radio energy. Rather we can only compute the fluence, which is the flux density integrated over the duration of the pulse. While we have not characterized the full spectrum, we do know that between 20 and 60 MHz the spectrum follows a power-law, given by $S \propto \nu^{\alpha}$, where $S$ is the flux density, $\nu$ is the frequency, and $\alpha$ is the spectral index \citep{Obenberger16}. For the 4 cases studied in \citet{Obenberger16}, $\alpha \sim -4\pm1$ during the peak of the afterglows. Therefore we scale the fluence of each afterglow presented here to 38 MHz using a spectral index of $-4$. We also scale the fluence to a distance of 100 km, which is the same scaling distance for optical absolute meteor magnitude. 

\begin{figure}
	\centering
	\includegraphics[width = 5.5in]{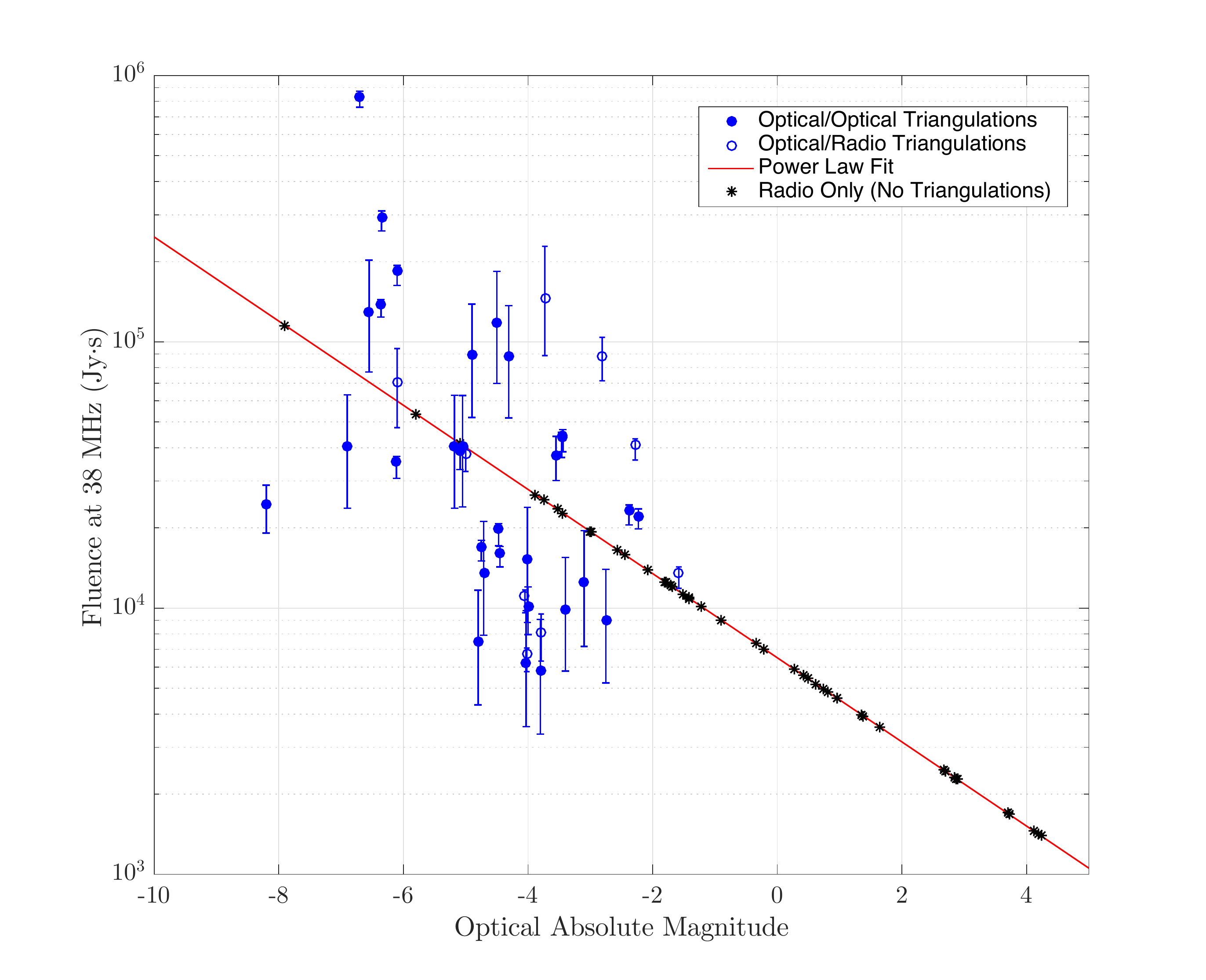}
	\caption{A plot showing the fluence at 38 MHz vs. peak absolute magnitude of the 40 radio afterglows with magnitude calibrated optical counterpart meteors, a power law fit, and the fluence of radio only (positions not triangulated) events plotted along the power law fit. Errors in the spectral index, position of the meteor,  beam pattern calibration, and image noise are included. Note: Spectral index error only affects observations at frequencies other than 38 MHz. Afterglows with optical/optical triangulations (solid) are shown separate from optical/radio triangulations (open). }
	\label{M_v_F}
\end{figure}

Figure \ref{M_v_F} compares the scaled fluence of the radio afterglow to the peak optical absolute meteor magnitude of the meteor counterpart along with a power law fit. There isn't a one to one correlation between the two observables, but there does appear to be a slight trend, where brighter optical events produces more radio energy. This relation may explain why roughly half of the radio afterglows did not have detected optical counterparts. Indeed, the detection rate of the ASGARD fireballs begins to drop for events below $-3$ \citep{Brown10}, and plotting the measured fluence of each optically undetected meteor along the power law shows that the majority would be dimmer than $-3$. While the large amount of scatter makes the fit meaningless, it does provide a reasonable explanation for the optically unobserved events.

There are many sources of error in the fluence calculation. First, there is a  systematic calibration error tied to the error in the assumed value of flux density of Cygnus A. This error is around 15\%, but since it would add a systematic offset, it is not shown in the plot. Sources of error that are shown in the plot include: error in the spectral index ($\pm 1$), errors in the position of the meteor (typically $\pm 4$ km), error in the beam pattern calibration ($\pm 5\%$), and image noise ($\pm 1 \sigma$). The spectral index error, which doesn't effect measurements taken at 38 MHz, is the dominant source of error.

\subsection{Altitudinal Dependence}

\begin{figure}
	\centering
	\includegraphics[width = 5.5in]{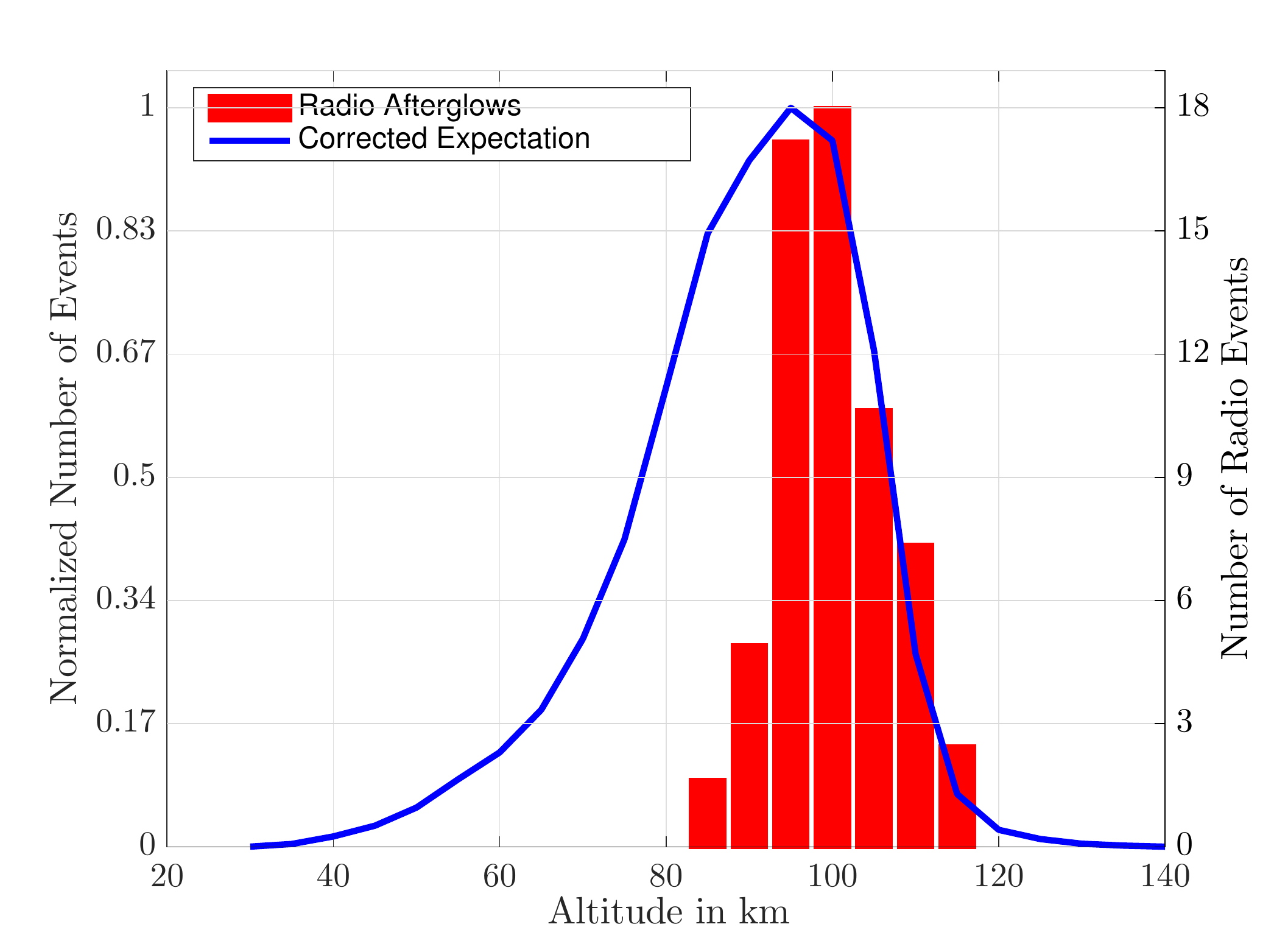}
	\caption{Histogram showing the expected number of bright meteors as a function of altitude above sea level (blue), relative to the maximum value, and the number measured of radio afterglows (red). The bin size of 5 km was used to match the typical amount of error in the position measurements. Above 90 km the measured distribution of radio events closely resembles the expected value, but below 90 km the number of radio events drops off much faster than what is expected. This suggests that there is a strong altitudinal dependence (perhaps even a strong cutoff) on radio emission excitation. The expected distribution is based on 1382 meteors observed between 2012 and 2013 from the two NASA/MEO cameras in Southern NM, and have been corrected for for the bias that arises from the fact that low altitude events will have a lower detection rate near the horizon.}
	\label{Altitude_Hist}
\end{figure}

Figure \ref{Altitude_Hist} shows the distribution of distinct radio components from the 44 triangulated radio afterglows as a function of altitude. Many events had several resolved radio components spread along the trail, allowing for multiple measurements each. Therefore the 44 triangulated afterglows resulted in 76 distinct component altitude measurements. The event shown in Figure \ref{image} was chosen to highlight an example of a afterglow with many individual radio components. This event is not typical; most radio afterglows are unresolved point sources as seen by the LWA1.

We compare the radio measurement distribution to the expected bright meteor altitude distribution, which is based on the tracks of 1382 bright meteors detected between January 1, 2012 and December 31, 2013 by the two NASA/MEO cameras at Las Cruces (LC) and Mayhill (MH) in southern New Mexico. This distribution represents the relative number of optical meteors (absolute magnitudes brighter than $\sim$ -2) that pass through each altitude bin.

We corrected the optical distribution for the bias that arises from the fact that low altitude events will have a lower detection rate near the horizon. This was corrected by randomly generating a large sample ($\sim 3\times10^{6}$) of latitudes, longitudes and elevations. We then calculated the azimuth and elevation angles of these positions as viewed from the LC and MH cameras. Those events that occurred below the visible horizon were excluded. This produced an elevation-dependent distribution, which we could use to back out a corrected distribution from the measured one. We then performed the same procedure using the mutual visibility of the MH camera and the LWA1. 

We used these corrections to predict the mutually observed distribution from MH and LWA1. The visible horizon is set by both the CCD cutoff to the north and south of the video images (see Fig. \ref{image}), as well as buildings, trees, mountains, etc., and the horizon is set to 25$^{\circ}$ elevation for the LWA1. We only correct the predicted bias between LWA1 and MH since the vast majority of events were detected by the MH camera. The correction resulted in very little change ($<$ 0.03 for any one bin), so we do not show the uncorrected version in Figure \ref{Altitude_Hist}.

The radio afterglow distribution matches well with the expected meteor distribution at the 95 km bin and higher, but below this the two distributions differ dramatically. While the expected distribution has a slow and steady power law drop off from 85 to 40 km, the detected radio distribution cuts off sharply below 90 km, with the lowest value being 85 km. This cutoff occurs despite the fact that many of the observed optical meteors reach altitudes and even peak in brightness far below this. Since the expected distribution traces the events that could be seen with the LWA1 with measurable altitudes, we conclude that the cutoff below 90 km is a feature inherent of radio afterglows. This finding suggests that in the 85 to 95 km range there exists some sort of barrier to the development of radio afterglows. 

An altitudinal dependence could also be consistent with a velocity dependence, due to higher velocity meteors ablating at higher altitudes \citep{Ceplecha98}. However, we note that 5 of the 31 optical/optical triangulated meteors had velocities less than 45 km/s, the slowest being 30 km/s. It is interesting to note that while these events did penetrate deeper into the atmosphere (down to 65 km), the radio afterglows were only seen at altitudes above 85 km. This was despite that fact that the brightest optical regions occurred at lower altitudes. 

\section{Discussion}

As mentioned in Section 1, plasma waves would have a lower altitudinal cutoff, because collisional damping would suppress wave growth. We have now measured a strong altitudinal dependence, which is potentially caused by this effect. 

From \citet{Schunk09} we can estimate the electron-ion ($\nu_{ei}$) and electron-neutral ($\nu_{en}$)  collision frequencies, where the neutral contribution is dominated by N$_2$, O$_2$, and O.  The yearly average number densities for N$_2$, O$_2$, and O and electron temperature (assuming equilibrium with surroundings) can be estimated using the Mass Spectrometer and Incoherent Scatter Model (MSIS) \citep{Picone02}. For ion collisions we assume single ionization, where the ion number density is the same as the electron number density.

Based on VHF overdense echo observations we know that plasma frequencies up to hundreds of MHz are possible. We can assume that the emission covers a large range of plasma frequencies, considering that a turbulent trail contains steep density gradients.

\begin{figure}
	\centering
	\includegraphics[width = 5.5in]{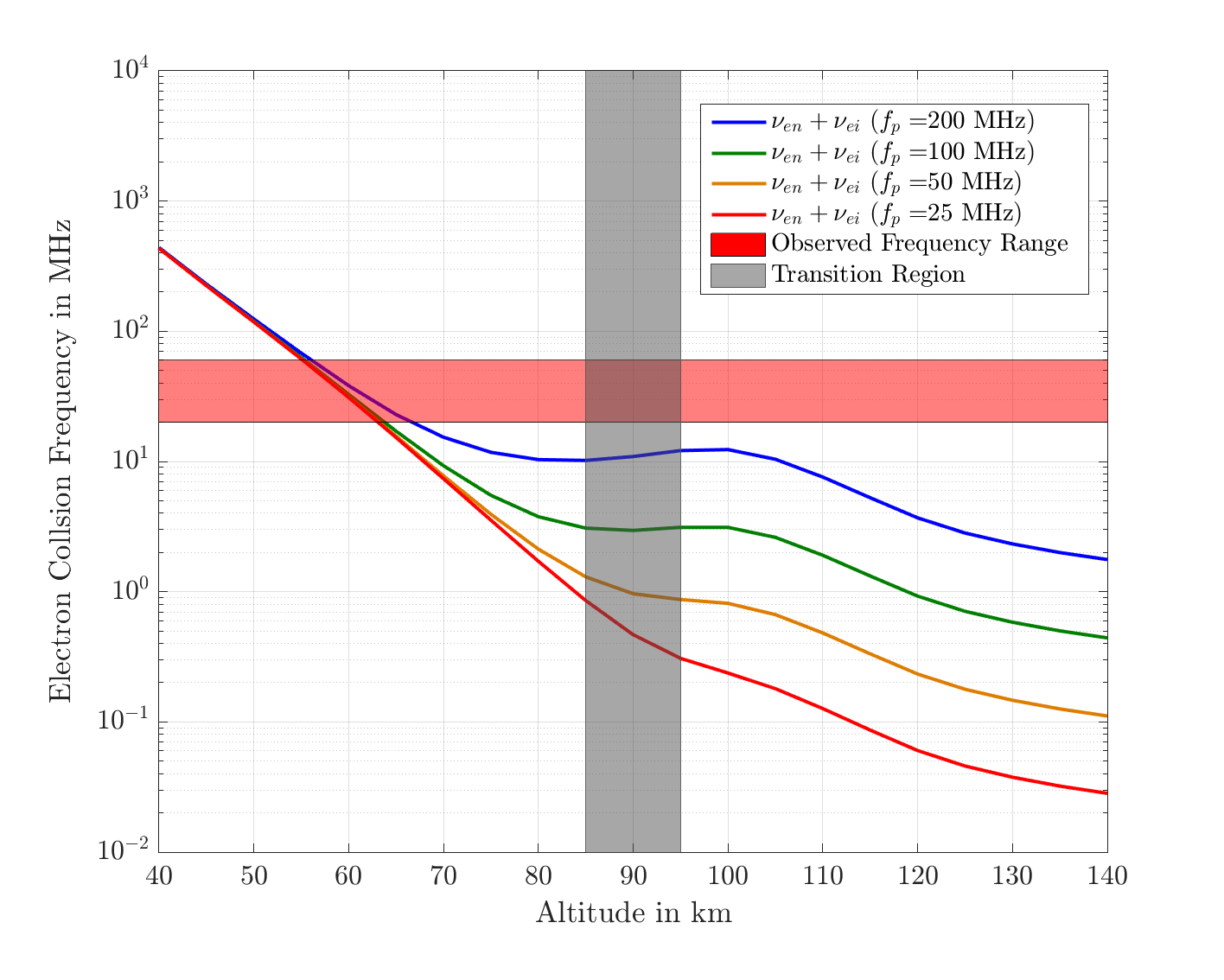}
	\caption{Plot of the estimated electron collision frequency (in MHz), as a function of altitude for plasma frequencies of 25, 50, 100, and 200 MHz, the measured transition region below which radio afterglows have not been observed (gray), and the frequency range where afterglows have been observed with the LWA1 (red).  }
	\label{collision}
\end{figure}

Figure \ref{collision} shows a range of collision frequencies as a function of altitude derived from different plasma frequencies, up to 200 MHz. Higher plasma frequencies correspond to large values of $n_{e}$ and therefore greater electron/ion collisions frequencies and damping. Neutral collisions dominate at lower altitudes ($<$ 70 km) and ion collisions dominate at higher altitudes ($>$ 100 km).

Based on collision frequency values shown in Figure \ref{collision}, we would expect a cutoff at an altitude of $\sim$ 65 km, where the collision frequency is on the same order of magnitude as the presumed plasma frequencies. However, depending on the strength of the wave driving mechanism, a lower altitude cutoff above 65 km may exist where the collisional damping rate is greater than the growth rate of the driving mechanism. Near this altitude would be a transitional zone where the collisional damping rate competes with the growth rate. The findings presented in this paper suggest this transition zone is around 85-95 km for afterglows between 25-38 MHz. Therefore, Figure \ref{collision} suggests a driving mechanism growth rate around $10^{6}$ s$^{-1}$. 

Though the findings presented in this paper agree with the plasma wave hypothesis, we note that there may indeed be some other reason why the afterglow emission has such a strong altitudinal dependence. For instance meteor trails show a variety of initial structures and evolutionary behaviors across altitudes. These factors may effect the radiation process in ways not yet considered. For instance, electromagnetic conversion of electron plasma waves could be hindered at lower altitudes.

As discussed in \citet{Obenberger14} and \citet{Obenberger15b} it is unlikely that meteor radio afterglows are caused by reflected broadband radio noise. However, overdense reflections would see an altitude dependent attenuation where lower altitude events contain more collisionaly induced absorption and less reflection \citep{Foschini99}. While there is no known source capable of producing such reflections, results presented in this paper could be consistent with such a phenomenon, which we have not been able to completely rule out.

Fortunately, we have recently completed construction of another full LWA station, 75 km northeast of LWA1, at the Sevilleta National Wildlife Refuge (LWA-SV). Combined LWA1 and LWA-SV observations of the same meteor afterglow will rule out any reflection process. Furthermore, this will remove the necessity for optical observations, since we will be able to triangulate the radio afterglows directly. While we corrected for geometrical bias of the optical cameras, we cannot be sure that our method of using optical cameras for triangulation did not contribute unconsidered sources of bias. Over the next few years we will use both LWA1 and LWA-SV to test the findings of this study. We will also be able to look for diurnal variations since we will be able to triangulate daytime afterglows. LWA-SV is currently in the commissioning phase and as of May 2016 is producing all-sky images.

\section{Conclusion}

Using coordinated optical and radio observations, we have found an altitude dependence on the occurrence of meteor radio afterglows. This dependence is characterized by a cutoff below 90 km. Such a cutoff agrees with our hypothesis that meteor radio afterglows are the result of electron plasma waves emission from turbulent meteor trails, though the cause may indeed be unrelated.

\begin{acknowledgments}
We thank the anonymous referees for thoughtful comments.

We thank Bjorn Gustavsson for helpful discussions and the use of the open source AIDA\_tools Matlab package for camera geometric calibrations (http://www.alis.irf.se/$\sim$bjorn/AIDA\_tools).

We thank Dwayne Free, Andy Howell, and Richard Spalding for providing and calibrating the Sky Sentinel data.

We thank Danielle Moser for useful discussions, and we thank both D. Moser and Bill Cooke for providing the NASA/MEO data.

This research was performed while the primary author, Kenneth Obenberger, held an NRC Research Associateship award at the Air Force Research Laboratory, Kirtland AFB, NM

Construction of the LWA1 has been supported by the Office of Naval Research under Contract N00014-07-C-0147. Support for operations and continuing development of the LWA1 is provided by the National Science Foundation under grants AST-1139963 and AST-1139974 of the University Radio Observatory program. 

All of the LWA1 data used in this article is publicly available at the LWA1 Data Archive (http://lda10g.alliance.unm.edu), optical data can be provided upon request.

\end{acknowledgments}

\end{article}

\end{document}